\documentclass{kapedbk}
\usepackage{graphicx,amsmath,amsfonts}
\normallatexbib

\begin{document}
\articletitle{Mesoscopic Aharonov-Bohm oscillations in metallic rings}

\author{T.~Ludwig}
\affil{Institut f\"ur Nanotechnologie,\\
  Forschungszentrum Karlsruhe, 76021 Karlsruhe, Germany}
\email{ludwig@int.fzk.de}

\and 

\author{A.~D.~Mirlin\footnote{Also at Petersburg Nuclear Physics
Institute, 188350 St.~Petersburg, Russia.}}
\affil{Institut f\"ur Nanotechnologie,\\
Forschungszentrum Karlsruhe, 76021 Karlsruhe, Germany\\ and\\
Institut f\"ur Theorie der Kondensierten Materie,\\
Universit\"at Karlsruhe, 76128 Karlsruhe, Germany}
\email{mirlin@tkm.physik.uni-karlsruhe.de}

\begin{abstract}
We study the amplitude of mesoscopic Aharonov-Bohm oscillations in
quasi-one-dimensional (Q1D) diffusive rings. We consider first the
low-temperature limit of a fully
coherent sample. The variance of oscillation harmonics is calculated
as a function of the length of the leads attaching the ring to reservoirs.
We further analyze the regime of relatively high temperatures, when
the dephasing due to electron-electron interaction suppresses
substantially the oscillations.
We show that the dephasing length $L_\phi^{\rm AB}$ governing the
damping factor $\exp(-2\pi R /L_\phi^{\rm AB})$ of the oscillations
is parametrically different from the common dephasing length for the
Q1D geometry. This is due to the fact that the dephasing is governed
by energy transfers determined by the ring circumference $2\pi R$,
making $L_\phi^{\rm AB}$ $R$-dependent.
\end{abstract}


\newcommand{\be}{\begin{equation}}
\newcommand{\ee}{\end{equation}}
\newcommand{\bea}{\begin{eqnarray}}
\newcommand{\eea}{\end{eqnarray}}
\newcommand{\br}{{\bf r}}

\section{Introduction}

The Aharonov-Bohm (AB) oscillations of conductance are one of the most
remarkable manifestations of electron phase coherence in mesoscopic
samples. Quantum interference of contributions of different electron
paths in a ring threaded by a magnetic flux $\Phi$ makes the
conductance $g$ an oscillatory function of $\Phi$, with a period
of the flux quantum $\Phi_0=hc/e$; see
Refs.~\cite{Washburn,Aronov_Sharvin,Imry} for reviews.
In a diffusive ring these $\Phi_0$-periodic conductance oscillations
are sample-specific (and would vanish upon the ensemble averaging),
due to a random phase associated with diffusive paths. In this respect,
the $\Phi_0$-periodic AB effect is a close relative of mesoscopic
conductance fluctuations. 

Another type of the AB effect is induced by interference of
time-reversed paths encircling the ring and is intimately related to
the weak localization (WL) correction \cite{Aronov_Sharvin}. Its principal
period is $\Phi_0/2\,$. It survives the ensemble averaging but is
suppressed by a magnetic field penetrating the sample. Below we
concentrate on the first (mesoscopic, or $\Phi_0$-periodic) AB
effect. Our results for the dephasing are, however, applicable to the second
(weak-localization, or $\Phi_0/2$-periodic) type of AB oscillations as
well, as we discuss in the end of Section~\ref{dephasing}.  

Interaction-induced inelastic processes lead to dephasing of
electrons, and thus to a damping of interference phenomena, in
particular of AB oscillations. The mesoscopic AB
oscillations can thus serve as a ``measuring device'' for the electron
decoherence. This idea was, in particular, implemented in recent
experiments \cite{Pierre_Birge_PRL,Birge_2003}, where the
low-temperature behavior of the dephasing time $\tau_\phi$ was
studied, and two mechanisms of decoherence were identified: scattering
off magnetic impurities and electron-electron scattering.

In Section~\ref{non-int} we will calculate the variance of harmonics
of mesoscopic Aharonov-Bohm oscillations in the low-temperature regime when
the dephasing effects are negligible.
In Section~\ref{dephasing} we will then analyze the opposite limit of
high temperatures when the dephasing due to electron-electron
interaction strongly suppresses the amplitude of the AB oscillations. 


\section{\label{non-int}Low-temperature limit: fully coherent sample}

In this section we will study the mesoscopic AB oscillations in the
low-temperature regime when both the thermal length $L_T$ and the
dephasing length $L_\phi^{AB}$ (see Sec.~\ref{dephasing}) are much
larger than the sample size. These conditions correspond to the regime
of universal conductance fluctuations \cite{Lee_Stone,Lee_Stone_Fukuyama,Kane_Serota_Lee}.
In this limit, the variance of conductance fluctuations of a Q1D wire
takes a universal value $8/15$ (in units of $(e^2/h)^2$) in the
absence of spin-orbit interaction, or $2/15$ for strong spin-orbit
interaction. If the time-reversal symmetry is broken by the magnetic
field, these values are reduced by a factor of $2$.

A natural question is what is the counterpart of these universal
values for the AB oscillations in a mesoscopic ring. It turns out,
however, that the situation in this case is more delicate, and the
amplitude of the oscillations depends in a non-trivial way on the
length of the wires connecting the ring to the bulk electrodes.

We will assume that the wires forming the ring are thin (i.e.~of Q1D
character), which allows us to solve the problem
analytically. Although the formalism we will use for this purpose is
well-known \cite{Lee_Stone,Lee_Stone_Fukuyama,Altshuler},
we are not aware of such a calculation in the literature. 
In some earlier papers, the problem of mesocopic AB oscillations was
studied numerically \cite{DiVincenzo_Kane,Mueller-Groeling}.
In Ref.~\cite{Loss_Schoeller_Goldbart} some analytical calculations
were performed, but the role of leads connecting the ring to
the reservoirs was completely disregarded. In paper~\cite{Falko} the
contacts were included, but only as an escape probability at the
junctions. The diffusive dynamics of electrons in the leads was not
taken into account. For this reason our results in this section, although qualitatively
similar to the results of \cite{Falko} in the no-dephasing regime, do 
differ quantitatively.

\begin{figure}[h!]
\begin{center}
\includegraphics[width=1.0\linewidth]{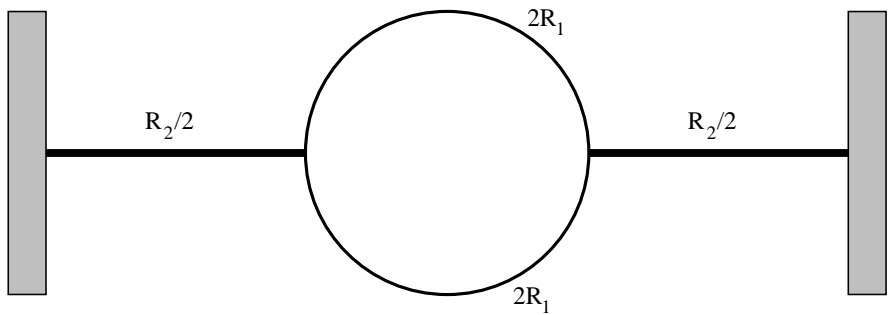}
\caption{\label{ring1}The sample geometry. The
  geometric parameter $\gamma$ is defined as the ratio of the
  resistance of the ring without the leads to the resistance of the
  total sample,
  $\gamma=R_1/(R_1+R_2)\,$.}
\end{center}
\end{figure}

We consider a thin (Q1D) ring coupled symmetrically by two leads to the bulk
electrodes. The only geometric parameter characterizing the
problem is then the ratio $\gamma$ of the resistance of the ring
itself to the total resistance of the ring with the leads (see Fig.~\ref{ring1}).
By definition, $0<\gamma<1$.

\subsection{Real-space formalism}

To calculate the conductance fluctuations of a ring with leads,
we use the formalism developed in
Refs.~\cite{Kane_Serota_Lee,DiVincenzo_Kane,Kane_Lee_DiVincenzo}.
To make the present paper self-contained we include a brief exposition
of the formalism.

The real-space DC conductivity $\sigma({\bf r},{\bf r^\prime})$ in the linear
response regime at zero temperature can be calculated by the Kubo
formula:
\be
\sigma_{\alpha\beta}(\br,\br^\prime) = \frac{e^2}{4\pi
  m^2}\,\nabla_\alpha\left[G^R(\br,\br^\prime)-G^A(\br,
    \br^\prime)\right] \nabla^\prime_\beta \left[G^A(\br^\prime,\br)-G^R(\br^\prime,\br)\right]
\label{kubo}
\ee
where $G^A$ and $G^R$ denote the advanced and retarded Green's
functions for electrons at the Fermi level. Note that by particle
number conservation the conductivity tensor must be divergenceless.

In real space, the impurity-averaged conductivity tensor $\langle\sigma\rangle$ is a long-ranged
object which in leading order can be expressed diagrammatically by the
sum of a bare conductivity bubble and a ladder diagram \cite{Kane_Serota_Lee}.
This sum is conveniently represented by defining a
``flow function'' $\phi$ (See Refs.~\cite{Kane_Serota_Lee,Kane_Lee_DiVincenzo} for details),
\be
\phi_{\alpha\beta}(\br,\br^\prime) \equiv
\delta_{\alpha\beta}\,\delta(\br-\br^\prime)-\nabla_\alpha\nabla_\beta^\prime{\cal
D}(\br,\br^\prime)
\ee
where ${\cal D}$ satisfies $-\nabla^2{\cal D}(\br,\br^\prime)=-\delta(\br,\br^\prime)\,$, so that
\be
\left\langle\sigma_{\alpha\beta}(\br,\br^\prime)\right\rangle =\sigma_0\,\phi_{\alpha\beta}(\br,\br^\prime)
\ee
where $\sigma_0=e^2\nu D$ is the Boltzmann conductivity.
Calculating the mesoscopic fluctuations of the conductivity (\ref{kubo}), one finds:
\bea
\lefteqn{\left\langle\delta\sigma_{\alpha\beta}(\br_1,\br_2)\,\delta\sigma_{\gamma\delta}(\br_3,\br_4)\right\rangle}\nonumber\\
&=&\int\!\!\!\int\!\!\!\int\!\!\!\int\! d\br_1^\prime d\br_2^\prime
d\br_3^\prime
d\br_4^\prime\,\phi_{\alpha\alpha^\prime}(\br_1,\br_1^\prime)\,\phi_{\beta\beta^\prime}(\br_2,\br_2^\prime)\,\phi_{\gamma\gamma^\prime}(\br_3,\br_3^\prime)\,\phi_{\delta\delta^\prime}(\br_4,\br_4^\prime)\nonumber\\
& &\hspace*{6cm}\times\,\Gamma_{\alpha^\prime\beta^\prime\gamma^\prime\delta^\prime}(\br_1^\prime,\br_2^\prime;\br_3^\prime,\br_4^\prime)\:.
\eea
Here $\Gamma$ is given by a set of two-diffuson and two-cooperon
diagrams \cite{Kane_Serota_Lee,Altshuler_Shklovskii}, yielding
\be
\Gamma_{xxxx}(\br_1,\br_2;\br_3,\br_4) =
24\,\delta(\br_1-\br_3)\,\delta(\br_2-\br_4)\,
\left|{\cal \tilde{P}}_D(\br_1,\br_2)\right|^2 \:,
\ee
where ${\cal \tilde{P}}_D$ is a rescaled diffusion propagator
satisfying $({-\rm i}\nabla-e{\bf A})^2\,{\cal
  \tilde{P}}_D(\br,\br^\prime)$\linebreak[2]$=$\linebreak[2]$-\delta(\br,\br^\prime)\,$.
To evaluate the conductance, the conductivity is integrated over the
cross sections of the leads. For a sample which consists of Q1D parts
we can switch to a one-dimensional formulation.
It is convenient to absorb a cross-sectional factor $S$ into the propagator,
so that the one-dimensional propagator ${\cal \tilde{P}}_D^{(1)}$ is defined as
\be
{\cal \tilde{P}}_D^{(1)}(x,x^\prime) \equiv S\,{\cal \tilde{P}}_D^{(3)}(\br,\br^\prime)\:,
\ee
where $S$ is the cross-section at the coordinate $x$.
The one-dimensional diffusion propagator
satisfies the diffusion equation $-\nabla^2 {\cal \tilde{P}}_D^{(1)}(x,x^\prime) =
\delta(x-x^\prime)\,$, and the conductance fluctuations are given by
\be
\label{dgdg}
\left\langle\delta g\,\delta g\right\rangle =
24\!\!\int\!\!dx_1^\prime\,\phi^2\left({\textstyle-\frac{L}{2}},x_1^\prime\right)\!\int\!\!dx_2^\prime\,
\phi^2\left({\textstyle \frac{L}{2}},x_2^\prime\right)\,{\cal \tilde{P}}_D(x_1^\prime,x_2^\prime)\,
{\cal \tilde{P}}_D(x_2^\prime,x_1^\prime) \:,
\ee
where $\pm\frac{L}{2}$ denote the ends of the leads at the reservoirs and
the system size $L$ is given by $L=\pi R/\gamma$, where $R$ is the
radius of the ring.

\subsection{Zero-temperature Aharonov-Bohm oscillations}
\label{zero-T}
It is convenient to work in a gauge where the vector potential gives
just a phase shift in the boundary conditions for the diffusion
propagator at the junctions. Then the Q1D diffusion propagator is a
linear function of each of the coordinates in each segment of the sample. The
coefficients are determined by the boundary conditions (unit jump in
the slope for equal coordinates, vanishing
propagators at the reservoirs, and particle number conservation --
modified by the phase shift due to the vector potential -- at the
junctions) and can be evaluated by solving a set of linear equations.
After a lengthy but straightforward calculation \cite{dipl},
integrating both coordinates of Eq.~(\ref{dgdg})
over the entire sample (including the leads) yields the correlator
\be
\left\langle \delta g(\Phi)\, \delta g(\Phi+\delta\Phi)\right\rangle = D(\delta\Phi)+C(2\Phi+\delta\Phi)\:,
\ee
where the diffuson contribution has the form
\bea
D(\delta\Phi)=
&\displaystyle\frac{1}{30}\Bigg[&(1-\gamma)^4
\,+\,\frac{320\,\gamma^2(1+\gamma)^4}{\left(1+6\gamma+\gamma^2-\left(1-\gamma\right)^2 {\rm cos}\left(2\pi\delta\Phi/\Phi_0\right)\right)^2}\nonumber\\
& &+\,\frac{16\,\gamma\,(1+\gamma)^2(1-10\gamma+\gamma^2)}{1+6\gamma+\gamma^2-\left(1-\gamma\right)^2
  {\rm cos}\left(2\pi\delta\Phi/\Phi_0\right)}\,\Bigg]\:.
\label{correlator}
\eea
The second term in (\ref{correlator}) is the cooperon contribution which has the same form as the diffuson
one with $\delta\Phi$ replaced by $2\Phi+\delta\Phi$ (i.e.~$C(x)=D(x)$) if the magnetic
flux threading the material of the ring is much less than one flux
quantum.
In the opposite limit the cooperon contribution is negligibly small.
Equation~\ref{correlator} corresponds to spinful electrons in the
absence of spin-orbit interaction; for strong spin-orbit interaction
the result is reduced by a factor of $4$.
Note that in Eq.~\ref{correlator} the limits of $\gamma\to 0$ (small
ring) and $\delta\Phi\to 0$ do not commute. If one first sets
$\gamma=0$ and then takes the limit $\delta\Phi\to 0$, one gets for
the correlation function the value $1/30\,$, whereas the opposite
order of limits yields the value $4/15\,$, as expected for a plain wire.

We define Fourier components of the conductance oscillations in the
following way:
\be
\delta g(\Phi) = \delta g_0+2\sum\limits_{n=1}^\infty \delta g_n\,{\rm
  cos}\left(2\pi n\Phi/\Phi_0+\theta_n\right)\:.
\label{e1}
\ee
The variance of the amplitude $\delta g_n$ of the $n$-th harmonic of
the oscillations is then found as the Fourier transform of
(\ref{correlator}), 
\bea
\left\langle \delta g_n^2\right\rangle&=&
\frac{1}{30}\left(\gamma^{1/2}-1\right)^{2n}\,\left(\gamma^{1/2}+1\right)^{-2n}\,\gamma^{1/2}\,(\gamma+1)\nonumber\\
& &\times\left[9-10\gamma+9\gamma^2+20n\,\gamma^{1/2}(\gamma+1)\right]
\label{n>0}
\eea
for $n\ge 1\,$, and
\be
\left\langle \delta g_0^2\right\rangle =
\frac{1}{30}\left(1+9\gamma^{1/2}-4\gamma-\gamma^{3/2}+6\gamma^2-\gamma^{5/2}-4\gamma^3+9\gamma^{7/2}+\gamma^4\right)\:.
\label{n=0}
\ee

\begin{figure}[h!]
\begin{center}
\includegraphics[width=1.0\linewidth]{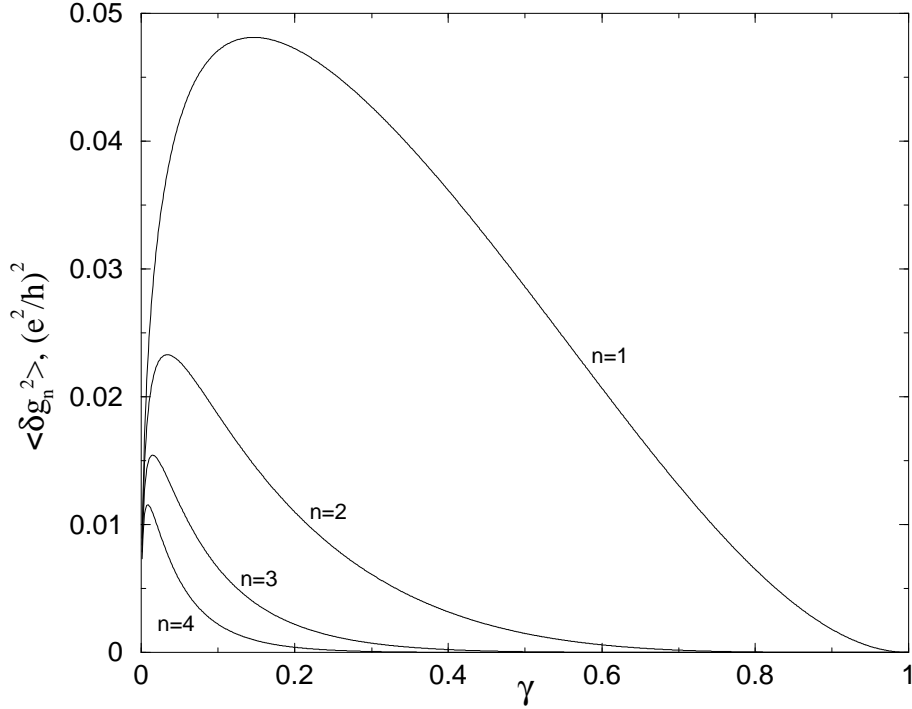}
\caption{\label{fourier-plot} Variance $\langle\delta g_n^2\rangle$ of
  the first four harmonics of the oscillations in the absence of
  spin-orbit interaction. With spin-orbit interaction,
  $\langle\delta g_n^2\rangle$ is reduced by a factor of~$4$.}
\end{center}
\end{figure}

The cooperon contribution does not change the oscillation amplitudes
Eq.~\ref{n>0}, but only affects the statistics of the phases $\theta_n\,$.
Specifically, if the magnetic flux through the material of the ring is
small and the cooperon contribution is present, $C(x)=D(x)\,$, the
phases $\theta_n$ are equal to $0$ or $\pi$. In the opposite limit,
when the cooperon contribution is suppressed,  the phases $\theta_n$
are randomized.
As to the aperiodic fluctuations, Eq.~(\ref{n=0}), the variance becomes
larger by a faCtor of $2$ is the presence of the cooperon term.

In Figure~\ref{fourier-plot} we plot $\langle\delta g_n^2\rangle$ for
$n=1,\ldots,4$ as a function of the geometric parameter $\gamma$.
It is seen that the oscillation amplitude depends on $\gamma$ in a
non-monotonous way, vanishing in both limits of long ($\gamma\to 0$)
and short ($\gamma\to 1$) leads.


\section{\label{dephasing}Dephasing by electron-electron interaction}

We now turn to the high-temperature regime and analyze how the
oscillations are suppressed due to dephasing induced by
electron-electron scattering processes. We will be mainly interested
in the exponential suppression factor and will not calculate the ($\gamma$-dependent)
numerical prefactor of order unity.

Within the conventional approach, when the dephasing time
$1/\tau_{\phi}$ is introduced as  a mass of the diffuson and cooperon
propagators, ${\cal P}_{D,C}(q,\omega)$\linebreak[2]$\sim$\linebreak[2] $ 1/(Dq^2-i\omega+1/\tau_\phi)$,
the variance of the $n$-th harmonic of the oscillations is suppressed
by the factor \cite{Aronov_Sharvin}
\be
\label{e2}
\left\langle \delta g_n^2\right\rangle \sim \frac{L_T^2\,L_\phi}{R^3} 
e^{-2\pi R n/L_\phi}\:,
\ee
where $L_\phi=(D\tau_\phi)^{1/2}$ is the dephasing length,
$L_T=(D/T)^{1/2}$ the thermal length, $D$ the diffusion constant, 
$T$ the temperature, $R$ the radius of the ring, and we set $\hbar=1$.
(It is assumed in Eq.~(\ref{e2}) that $L_\phi,L_T\ll 2\pi R$.)
For a thin ring, $L_\phi$ is then
identified with the dephasing length governing the WL 
correction in the quasi-one-dimensional (Q1D) geometry, which was
found by Altshuler, Aronov and Khmelnitskii \cite{AAK,Altshuler_Aronov} to be 
\be
\label{e3}
L_\phi=\left(D\tau_\phi\right)^{1/2},\qquad \tau_{\phi}^{-1}\sim 
\left(\frac{T}{\nu D^{1/2}}\right)^{2/3}\:.
\ee
In fact, Aleiner and Blanter showed recently \cite{Aleiner_Blanter}
that $\tau_\phi$ relevant to the mesoscopic conductance fluctuations in
wires has indeed the same form, Eq.~(\ref{e3}),
as the WL dephasing time. 
This seems to support the assumption that the dephasing times
governing different mesoscopic phenomena are identical. Equations
(\ref{e2}), (\ref{e3}) are commonly used for the
analysis of experimental data. 

We will show below, however, that contrary to the naive expectations
the formulas (\ref{e2}) and
(\ref{e3}) do not describe correctly the dephasing of AB
oscillations. Specifically, if the interaction-induced exponential
damping factor of AB oscillations is presented in the form 
$\langle\delta g_n^2\rangle\sim\exp(-2\pi R n/L_\phi^{\rm AB})$,
the corresponding length $L_\phi^{\rm AB}$ is parametrically
different from Eq.~(\ref{e3}). Moreover, $L_\phi^{\rm AB}$ depends on the
system size $R$. 

\subsection{Effective electron-electron interaction}

Following Refs.~\cite{AAK,Altshuler_Aronov,Aleiner_Blanter,AAG}, the
elec\-tron-\-elec\-tron interaction can be represented 
 by external
time-dependent random fields $\varphi^\alpha(\br,t)\,$, 
with the correlation function 
$\langle\varphi^\alpha(\br,t)\,\varphi^\beta(\br',t')\rangle$
determined from the fluctuation-dissipation theorem,
\be
\label{e4}
\left\langle\varphi^\alpha({\bf r})\,\varphi^\beta({\bf
r^\prime})\right\rangle_\omega = -{\rm Im}\,U({\bf r},{\bf r^\prime;\omega})\,
\delta_{\alpha\beta}\,{\rm coth}\frac{\omega}{2T}\:.
\ee
The conventional form for the dynamically screened Coulomb interaction
in a diffusive system is \cite{Altshuler_Aronov} 
\be
\label{e5}
U(q,\omega)=\frac{1}{U_0^{-1}(q)+\Pi(q,\omega)}\simeq \Pi^{-1}(q,\omega)\:,
\ee
where $U_0(q)$ is the bare Coulomb interaction, 
$\Pi(q,\omega)= \nu Dq^2/(Dq^2-i\omega)$ is the polarization operator,
and $\nu$ is the density of states. As we will see below, the
characteristic momenta $q$ for our problem are of the order of the
inverse system size $R^{-1}$. In view of the non-trivial geometry of
our system, it is thus more appropriate to work in the coordinate
representation. A corresponding generalization of
Eq.~(\ref{e5}) can be readily obtained, yielding
\be
\label{e6}
{\rm Im}\,U(\br,\br';\omega)\simeq {\rm Im}\,
\Pi^{-1}(\br,\br';\omega)=-\frac{\omega}{\nu D}{\cal D}(\br,\br')\:,
\ee
where ${\cal D}$ is the propagator for the Laplace equation, 
$-\nabla^2{\cal D}(\br,\br^\prime)=\delta(\br-\br^\prime)\,$ with zero
boundary conditions at the contacts with bulk
electrodes.  Substituting Eq.~(\ref{e6}) in Eq.~(\ref{e4}), we get, for
relevant frequencies $\omega\ll T$,
\be
\label{e7}
\left\langle\varphi^\alpha(\br,t)\,\varphi^\beta(\br^\prime,t^\prime)\right\rangle = 
\frac{2T}{\nu D}\,{\cal
  D}(\br,\br^\prime)\,\delta_{\alpha\beta}\,\delta(t-t^\prime)\:.
\ee

\subsection{Aharonov-Bohm oscillations in presence of interaction}

Since the ring we are considering consists of Q1D wires, the
corresponding diffusion propagator satisfies the one-dimensional
diffusion equation
\bea
\label{e8}
\left\{\partial_t-D\partial_x^2+{\rm
  i}\left[\varphi^\alpha(x,t)-\varphi^\beta(x,t)\right]\right\}
\,{\cal P}_{\delta\Phi}^{\alpha\beta}(x,t;x',t')\nonumber\\ 
= \delta(x-x')\,\delta(t-t')
\eea
supplemented by appropriate matching conditions at junctions of the
ring and leads. Here $\delta\Phi=\Phi_1-\Phi_2$ is difference
in the AB flux between the two measurements, which is incorporated in
the matching conditions. The conductance correlation function in the
high-$T$ limit is again given by the conventional two-diffuson diagrams
\cite{Altshuler_Shklovskii,Kane_Serota_Lee} (again we drop the cooperon
contribution which affects only the phases but not the amplitudes of
the oscillations), yielding (we drop the prefactor of order unity)
\bea
\label{e9}
&& \big\langle\delta g(\Phi_1)\,\delta g(\Phi_2)\big\rangle \sim 
\frac{D^2}{TR^4} \int dx_1\,dx_2 \int dt\,
dt^\prime \nonumber\\ && \qquad
\times\,\tilde{\delta}(t-t^\prime)\,\big\langle
{\cal P}^{12}_{\delta\Phi}(x_1,x_2,t)\,
{\cal P}^{21}_{\delta\Phi}(x_2,x_1,t^\prime)
\big\rangle
\eea
where angular brackets denote averaging over the external fields,
$\tilde{\delta}(t)$ is given by
\bea
\label{e10}
\tilde{\delta}(t)&=&12\pi T\int\frac{d\epsilon_1}{2\pi}
\frac{d\epsilon_2}{2\pi}\,f'(\epsilon_1)\,f'(\epsilon_2)\,
{\rm e}^{{\rm i}(\epsilon_1-\epsilon_2)t}\nonumber\\
&=&3\,\pi\,T^3\,t^2 \,{\rm sinh}^{-2}\left(\pi Tt\right)\:,
\eea
and $f(\epsilon)$ is the Fermi distribution function. The function
$\tilde{\delta}(t)$ is peaked at $t=0$ with a width $T^{-1}$. We will
replace it below by the delta-function $\delta(t)$. This is justified
if the dephasing effect during the time $t-t'\sim T^{-1}$ is
negligible, i.e. $\langle \phi^\alpha(x,t)\phi^\alpha(x,t)\rangle
T^{-1} \ll 1$. Using Eq.~(\ref{e7}), we find that the latter condition is
equivalent to the requirement that the conductance of the sample is
much larger than the
conductance quantum $e^2/h \simeq (25\,{\rm k}\Omega)^{-1}$.
This condition is well satisfied in typical experiments with metallic
rings, thus justifying the replacement
$\tilde{\delta}(t)\to\delta(t)$. 

We now express the diffusion propagators in Eq.~(\ref{e9}) as path
integrals. We are interested in the regime of strong dephasing, when
the relevant paths propagate only inside the ring and do not extend
into the leads (see below). It is convenient to introduce the angular
coordinate $\theta$ on the ring ($-\pi\le\theta\le\pi$), with
$\theta=\pm\pi/2$ corresponding to the junctions with the leads.

\begin{figure}[h!]
\begin{center}
\includegraphics[width=1.0\linewidth]{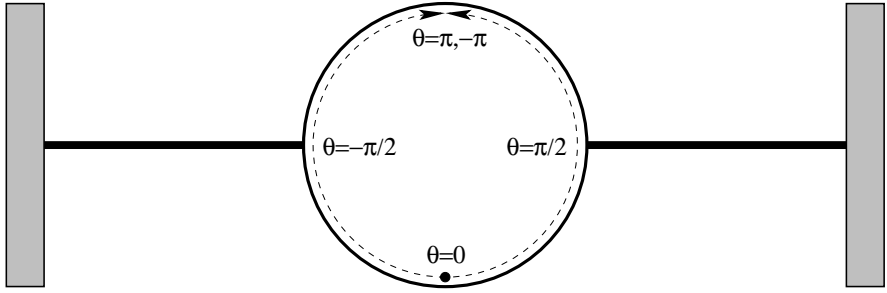}
\caption{\label{ring} The angular coordinate $\theta$ introduced
  above. The paths representing the saddle-point solution are shown.}
\end{center}
\end{figure}

Expanding the
conductance fluctuations in Fourier harmonics with respect to the
flux, $\delta g(\Phi)\to\delta g_n$, we then get
\bea
\lefteqn{\left\langle\delta g_n^2\right\rangle \sim \frac{D^2}{TR^4} 
\int d\Theta_1\,d\Theta_2 \int dt\,
\int\limits_{\Theta_2}^{\Theta_1}D\theta_1(t)
\int\limits_{\Theta_2}^{\Theta_1}D\theta_2(t)}& &\nonumber\\
&\times&\!{\rm exp}\left\{-\int\limits_0^t
dt^\prime\left[\frac{R^2\dot{\theta_1}^2}{4D}+\frac{R^2\dot{\theta_2}^2}{4D}+V(\theta_1,\theta_2)
\right]\right\},\:\:\:\:\:\:
\label{e11}
\eea
where the path integral  
goes over pairs of paths $\theta_1(t), \theta_2(t)$ which have a
relative winding number $n$. 
The ``potential'' $V(\theta_1,\theta_2)$ in Eq.~(\ref{e11}) is given by 
$V(\theta_1,\theta_2)=\langle(\phi^\alpha(\theta_1)
-\phi^\alpha(\theta_2))^2\rangle$; its explicit form can be
straightforwardly obtained according to Eq.~(\ref{e7}) by solving the
diffusion equation in the ring with leads as in Section~\ref{zero-T} and
presented in detail in \cite{dipl}.
We will only need below the form of $V(\theta_1,\theta_2)$ for both
coordinates being in the same arm of the ring. For
$|\theta_i| \le \pi/2$ we find
\be
\label{e12}
V(\theta_1,\theta_2)=\frac{2TR}{\nu D}
\left[\left|\theta_1-\theta_2\right|-\frac{\gamma+1}{2\pi}
\left(\theta_1-\theta_2\right)^2\right];
\ee
the expression for $|\theta_i|>\pi/2\,$ follows from symmetry
considerations.

\subsection{Strong-dephasing limit}

We consider first the fundamental harmonic ($n=1$) of the AB
oscillations; a generalization to higher harmonics, $n=2,3,\ldots$
will be done in the end. For $n=1$ the relevant pairs of paths
interfere after half-encircling the ring in the opposite directions. 
We are interested in the regime of a relatively high temperature, when
the dephasing effect is strong.
In this case, the path integral in Eq.~(\ref{e11})
can be evaluated via the saddle-point method. As has been mentioned
above, the paths representing the saddle-point solution (instanton) do
not extend into the leads. Indeed, exploring a part of a lead and
returning back into the ring would only increase the action 
of the path.
It is clear from the symmetry considerations that the optimal paths
satisfy $\theta_1(t)=-\theta_2(t)$. Furthermore, it is easy to see
that the optimal initial and final points are located at maximum
distance from the leads, i.e.~$\Theta_1=0$ and $\Theta_2=\pi$ (or vice
versa).
To within exponential accuracy, the problem is then reduced to that
of a particle of mass $R^2/D$ tunneling with zero energy in the
potential  
\be
\label{e13}
V(\theta)=\frac{4TR}{\nu D}\times \left\{
\begin{array}{ll}
\left[\theta-\frac{\gamma+1}{\pi}\theta^2\right]\,,\ \ 
& 0\le\theta\le\frac{\pi}{2} \\
\left[(\pi-\theta)-\frac{\gamma+1}{\pi}(\pi-\theta)^2\right]\,,\ \ 
&\frac{\pi}{2}\le\theta\le\pi
\end{array} \right.
\ee
from $\theta=0$ to $\theta=\pi$. Since the potential is composed of
quadratic parts, the corresponding instanton action is easily
calculated, yielding $\langle \delta g_1^2\rangle \propto e^{-S}$ with 
\be
\label{e14}
S = C_\gamma\,\frac{T^{1/2}\,R^{3/2}}{\nu^{1/2}\,D}\:.
\ee
Here $C_\gamma$ is a coefficient of order unity depending on
the geometrical factor
$\gamma$,
\be
\label{e15}
C_\gamma= \left[\frac{\pi}{2(\gamma+1)}\right]^{3/2}
\left[2\gamma\left(1-\gamma^2\right)^{1/2}
+\pi+2\,\arcsin\gamma\right];
\ee
$C_\gamma$ is equal to $\frac{\pi^{5/2}}{2^{3/2}}$ in the limit of long leads 
($\gamma\to 0$) and to $\frac{\pi^{5/2}}{4}$ in the limit of short leads 
($\gamma\to 1$).

\begin{figure}[h!]
\begin{center}
\includegraphics[width=1.0\linewidth]{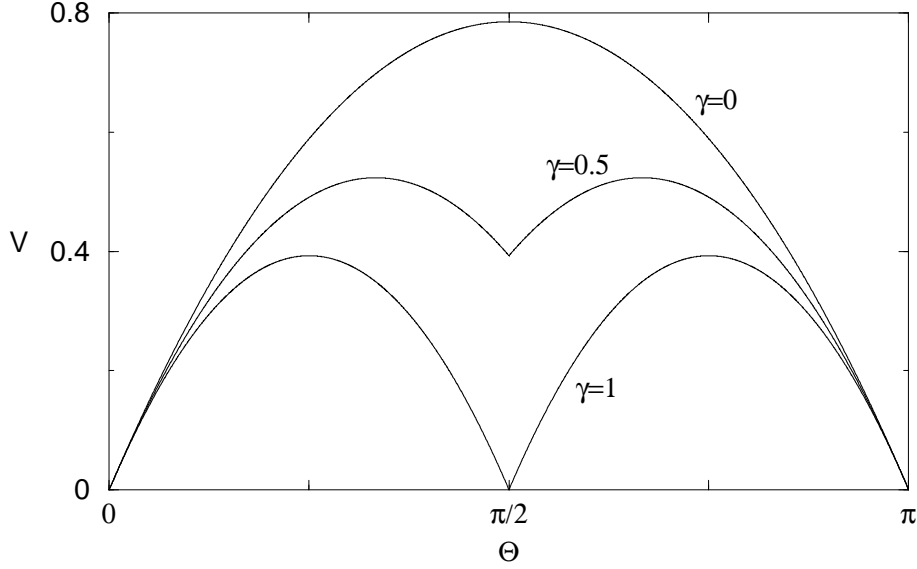}
\caption{\label{potential} The potential $V(\theta)$ of the tunneling
problem, Eq.~\ref{e13}, plotted in units of $4TR/\nu D$ for
different values of the geometric coefficient $\gamma$.}
\end{center}
\end{figure}

The above calculation can be straightforwardly
generalized to higher harmonics of the AB oscillations,
$n=2,3,\ldots\:$. The optimal paths still begin at $\Theta=0$ or $\pi$
but now perform $n/2$ windings in the opposite directions. Therefore,
the corresponding action is $S_n=nS$.

To calculate the preexponential factor, we have to take into account
small fluctuations of the initial and final points $\Theta_1,
\Theta_2$ around their optimal
values, as well as fluctuations of the paths $\theta_1(t),\theta_2(t)$
around the instanton solution.
 We will only calculate the
parametric dependence of the prefactor, neglecting numerical factors
of order unity.
First, let us consider small offsets of the initial and final
points of the paths from their optimal position. The second order
variation of the 
action $\delta^2S$ will be a quadratic form of the offsets, $\delta^2
S=u_{ij}\,\delta\Theta_i\,\delta\Theta_j$, where $i=1,2$. Using that 
$\delta^2S \sim 1$ for $\delta\Theta_i\sim 1$, we get 
\be
\left({\rm det}\,u_{ij}\right)^{-1/2} \sim S^{-1}\:.
\label{a1}
\ee
Second, we have to account for small deviations of the paths
from the instanton solution. The corresponding factor
can be identified as the propagator for a harmonic oscillator with the
parameters $m\sim R^2/D$ and
$m\omega^2\sim RT/D\nu$. There are two such factors
(one for each of the paths), yielding together
\be
\left[\left(m\omega\right)^{1/2}\right]^2\sim
\frac{T^{1/2}\,R^{3/2}}{\nu^{1/2}\,D}. 
\label{a2}
\ee
Finally, there is a Gaussian integration over the deviations of the time
$t$ spent on the path from its optimal value $t_{\rm opt}\sim (\nu
R/T)^{1/2}$. The corresponding factor can be estimated as 
\be
\label{a3}
\left(\frac{\partial^2S}{\partial t^2}\right)^{-1/2}_{t=t_{\rm
opt}} \sim \left(\frac{S}{t_{\rm opt}^2}\right)^{-1/2} 
\sim \frac{\nu^{3/4}\,D^{1/2}}{T^{3/4}\,R^{1/4}}.
\ee
Combining Eqs.~(\ref{e11}), (\ref{a1}), (\ref{a2}) and (\ref{a3}), 
we obtain the final result for the variance of the
harmonics  of the mesoscopic  AB oscillations
\be
\label{e16}
\left\langle\delta g_n^2\right\rangle \sim
\left(\frac{L_T}{R}\right)^{7/2}
\left(\frac{\nu D}{R}\right)^{3/4}{\rm e}^{-nS}
\end{equation}
where $n=1,2,\ldots\:$, 
and the action $S$ is given by Eq.~(\ref{e14}). 

\subsection{Aharonov-Bohm dephasing time}

Let us discuss the obtained result (\ref{e16}), (\ref{e14}). 
First of all, it is
essentially different from what one would obtain by using the formulas
(\ref{e2}), (\ref{e3}). Indeed, the exponent in Eq.~(\ref{e16}) scales in a
different way with the temperature and with the system size, as
compared to Eqs.~(\ref{e2}), (\ref{e3}). It is instructive to rewrite
Eq.~(\ref{e16}) in a form analogous to Eq.~(\ref{e2}),
\be
\label{e17}
\left\langle\delta g_n^2\right\rangle \sim \left(\frac{L_T}{R}\right)^2
\left(\frac{L_\phi^{\rm AB}}{R}\right)^{3/2}
e^{-2\pi n R/L_\phi^{\rm AB}}\:, 
\ee
thus defining the Aharonov-Bohm dephasing length 
$L_\phi^{\rm AB}$,
\be
\label{e18}
L_\phi^{\rm AB}=\frac{2\pi}{C_\gamma}\,\frac{\nu^{1/2}\,D}{T^{1/2}\,R^{1/2}}\:.
\ee
The corresponding dephasing rate
$1/\tau_\phi^{\rm AB}=D/(L_\phi^{\rm AB})^2$  is thus given by
\be
\label{e19}
1/\tau_\phi^{\rm AB} = \left(\frac{C_\gamma}{2\pi}\right)^2\,
\frac{TR}{\nu D}\:.
\ee
To shed more light on the physical reason for the difference between
the conventional Q1D formula (\ref{e3}) and our result
(\ref{e18}), (\ref{e19}),
the following qualitative argument is instructive. 
Calculating perturbatively the dephasing rate using the formula
(\ref{e5}) for the screened Coulomb interaction in a diffusive system,
one gets
\be
\label{e20}
\tau_\phi^{-1}=\int \frac{dq}{2\pi}\,\frac{T}{\nu Dq^2}\:.
\ee
In the calculation of the dephasing rate in a wire 
\cite{Altshuler_Aronov,AAG,Aleiner_Blanter}, the arising infrared
divergence is cut off self-consistently, since only processes with
energy transfers $\omega >\tau_\phi^{-1}$ contribute. 
As a result, the
lower limit of integration in Eq.~(\ref{e20}) is $\sim L_\phi^{-1}$,
yielding the result (\ref{e3}). On the other hand, in the case of the
Aharonov-Bohm dephasing rate, the relevant paths have to encircle the
ring. Therefore, despite the fact that $L_\phi^{\rm AB}\ll 2\pi R$,
the low-momentum cutoff in Eq.~(\ref{e20}) is set by the inverse system
size $(2\pi R)^{-1}$. This yields $1/\tau_\phi^{\rm AB}\sim TR/\nu
D$, reproducing (up to a numerical coefficient) the result (\ref{e19}). 

It is worth emphasizing that our result Eq.~(\ref{e19}) for the dephasing
rate depends not only on the ring radius, but also on the geometry of
the leads through the coefficient $C_\gamma\,$. We note a certain
similarity between this result and the dependence of the dephasing
rate in a {\it ballistic} AB-ring on the probe configuration recently
found in \cite{Buttiker}.

As has been mentioned in the introduction, our results are also
applicable to the WL ($h/2e$-periodic) AB-oscillations. Their $n$-th
harmonic $\delta g_n^{\rm WL}$ is determined by cooperon paths with winding
number $n$. Assuming that the magnetic flux penetrating the sample is
negligible and comparing the path-integral representations for
$\langle\delta g_n^2\rangle$ and $\delta g_n^{\rm WL}\,$ we find
\be
\label{e21}
\left\langle\delta g_n^2\right\rangle = \frac{e^2D}{3TL^2}\,\delta
g_n^{\rm WL}
\ee
where $L=\pi R/\gamma$. This implies, in particular, that the dephasing
length $L_\phi^{\rm AB}\,$, Eq.~(\ref{e18}), is the same for both types
of the AB effect. Equation (\ref{e21}) is a generalization of the
relation \cite{Aleiner_Blanter} between the WL correction and
conductance fluctuations for single-connected geometries.


\section{Summary}

In Section~\ref{non-int} we have studied the amplitude of mesoscopic
Aharonov-Bohm oscillations in the low-temperature (phase-coherent)
regime. We have shown that, in contrast to fluctuations in a wire, the
oscillation amplitude does not take a universal value, but depends on
the geometric factor $\gamma$ characterizing the ratio of resistances
of the ring and the leads. The $\gamma$-dependence of the variance of
the oscillation harmonics is non-monotonous as shown in
Figure~\ref{fourier-plot}. For a typical experiment
\cite{Haeussler_Scheer_Weber_vLoehneysen,Strunk,Pierre_Birge_PRL},
where the resistances of the ring and the leads are comparable (so that
$\gamma\approx 0.5$), our results predict the r.m.s.~amplitude of the
first hamonic ${\rm rms}(2\,\delta g_1)\approx 0.16\,$. This value
compares well with the result of \cite{Strunk} at low bias voltage, as
well as with the low-temperature results of \cite{Pierre_Birge_PRL} in strong
magnetic fields, when the magnetic impurities are frozen.

In Section~\ref{dephasing} we have studied how the Aharonov-Bohm
oscillations are suppressed by dephasing caused by the
electron-electron interaction. Using the path integral formalism and
the instanton method, we have obtained the result (\ref{e16}),
(\ref{e14}) which is parametrically different from the naive
expectation (\ref{e2}), (\ref{e3}). This demonstrates that the
Aharonov-Bohm  dephasing rate $1/\tau_\phi^{\rm AB}$,
Eq.~(\ref{e19}),  is parametrically different from the dephasing rate
$1/\tau_\phi\,$, Eq.~(\ref{e3}), corresponding to a single-connected
geometry. Physically, the difference can be traced back to the fact
that $1/\tau_\phi$ is determined self-consistently by the processes
with energy transfers of the order of $1/\tau_\phi$ itself
(or equivalently with momentum transfers $\sim 1/L_\phi$),
while the characteristic energy and momentum transfers governing
$1/\tau_\phi^{\rm AB}$ 
are determined by the system size. For this reason, the Aharonov-Bohm
dephasing rate $1/\tau_\phi^{\rm AB}$ depends on the ring radius $R$,
diverging in the limit $R\to\infty\,$.

\begin{acknowledgments}
Valuable discussions with I.L.~Aleiner, B.L.~Altshuler,
N.O.~Birge, H.~Bouchiat, V.I.~Falko, I.V.~Gornyi and C.~Strunk 
are gratefully acknowledged.
We also thank M.~B\"uttiker for attracting our attention to
Ref.~\cite{Buttiker}.
\end{acknowledgments}

\end{document}